**Benchmarking of spin-orbit torque switching efficiency in Pt alloys**

*Chen-Yu Hu, Chi-Feng Pai**

Chen-Yu Hu, Prof. Chi-Feng Pai

Department of Materials Science and Engineering

National Taiwan University

Taipei 10617, Taiwan

E-mail: cfpai@ntu.edu.tw

## Abstract

A magnetic heterostructure with good thermal stability, large damping-like spin-obit torque (DL-SOT) and low power consumption is crucial to realize thermally stable, fast and efficient magnetization manipulation in SOT devices. This work systematically investigates on $Pt_xCu_{1-x}$/Co/MgO magnetic heterostructures with perpendicular magnetic anisotropy (PMA), and reports a promising spin Hall material, Pt-Cu alloy, possessing large DL-SOT efficiency and moderate resistivity. The optimal $Pt_{0.57}Cu_{0.43}$ has a large DL-SOT efficiency about 0.44, as determined by hysteresis loop shift measurements, with a relatively low resistivity (82.5 μΩ·cm at 5 nm thickness). Moreover, this large DL-SOT efficiency and the coercivity reduction accompanying with proper alloying contribute to a low critical switching current density ($2.37\times 10^6$ A·cm$^{-2}$ in the $Pt_{0.57}Cu_{0.43}$ layer) in current-induced magnetization switching measurements. Finally, the thermal stability of the Co layer can be preserved under alloying, whereas the switching power consumption can be significantly reduced, being the best performance among reported Pt-based spin current sources. Our systematic study on SOT switching properties suggest that $Pt_{0.57}Cu_{0.43}$ is an attractive spin current source with moderate resistivity, large DL-SOT efficiency, good thermal stability and low power consumption for future SOT applications.

## 1. Introduction

The spin-orbit torques (SOTs) exerted by spin currents ($J_s$) on ferromagnetic (FM) layer is capable to effectively manipulate magnetization in nanoscale, such as magnetization switching[1-3], magnetic domain wall motion[4, 5], magnetization dynamics at microwave or terahertz frequency[6]. Among versatile applications, SOT magnetic random-access memories (SOT-MRAM) promises merits, such as energy efficiency, fast operation speed, and therefore is an attractive candidate to replace the contemporary memory technologies. Unlike the spin-polarized $J_s$ in spin-transfer torque (STT)-MRAM, the pure spin current $J_s$ in SOT-MRAM majorly arises from bulk spin Hall effect (SHE)[7, 8], interfacial Rashba effect[9, 10] or spin-momentum locking[11-13]. The SHE is generally originated from 5d heavy metal (HM) with strong spin-orbit coupling (SOC), such as Pt[14], Ta[1], W[15]; The interfacial Rashba effect happens at the interface with strong SOC and broken inversion symmetry, such as Bi/Ag[16], α-Sn/Ag[17], $Bi_2Se_3$/Ag[18] interfaces; The spin-momentum locking stems from the topologically protective surface states of topological insulators, such as BiSe[19-21], BiSb[22], $(BiSb)_2Te_3$[23].

For SOT-MRAM applications, the most critical figures of merit are scalability, thermal stability, and power consumption. The first two factors can be optimized by introducing FM layer with decent perpendicular magnetic anisotropy (PMA)[24]. Common approaches include employing HM/FM interface with strong SOC[25], orbital hybridization at the FM/oxide interface[26, 27], or bulk PMA materials[28]. As for power consumption, the power required for reversing a single bit per volume, excluding current shunting to the FM layer is $p \propto \rho_{\mathrm{HM}} J_{\mathrm{HM}}^2$, where $\rho_{\mathrm{HM}}$ is the longitudinal resistivity of heavy metal, and $J_{\mathrm{HM}}$ is the switching charge current density in the heavy metal channel. Generally, the SOT switching is governed by the damping-like SOT (DL-SOT), and the DL-SOT efficiency ($\xi_{DL}$) is inversely proportional to the critical $J_{\mathrm{HM}}$. Under the SHE scenario, the $\xi_{DL}$ with a perfect spin transparency ($T_{int}$) at the HM/FM interface can be written as[29]

$$\xi_{DL} \equiv \left(\frac{2e}{\hbar}\right)\sigma_{SH}^{\mathrm{HM}}\rho_{xx}^{\mathrm{HM}}, \tag{1}$$

where $e$ is the elementary charge; $\hbar$ is the reduced Plank constant; $\sigma_{SH}^{\mathrm{HM}}$ and $\rho_{xx}^{\mathrm{HM}}$ respectively are spin Hall conductivity and resistivity of HM. Besides improving power consumption, lower current density can also avoid device degradation due to electromigration and then ensure the device endurance, a practical scale is below few $10^7$ A·cm$^{-2}$. To reduce $J_{\mathrm{HM}}$, a straightforward approach is to enhance $\xi_{DL}$ by using more resistive spin current source (β-W[30], β-Ta[1], BiSe[19], BiSb[22]) or raising $\rho_{xx}^{\mathrm{HM}}$ in certain materials system (nitrogen/oxygen doping[31], crystallinity engineering[32], alloying[33-37], thin layer insertion[38, 39]), based on the theoretical prediction[40] that intrinsic $\sigma_{SH}^{\mathrm{HM}}$ is independent of resistivity change. So far, the reported low $J_{\mathrm{HM}}$ ranges between $10^6$ to $10^7$ A·cm$^{-2}$. However, with raising $\rho_{xx}^{\mathrm{HM}}$, possible side-effects include the increase of power consumption per switching, energy dissipation due to detrimental current shunting, or poor SOT device endurance[41]. Moreover, engineering on spin Hall metal could possibly weaken the PMA of FM layer, thereby attenuate the device retention. As a result, it is crucial to find an optimized material system simultaneously possessing good PMA, large DL-SOT efficiency and moderate resistivity for a more applicable SOT-MRAM.

In this work, we choose co-sputtered Pt-Cu alloy as the tentative spin current source. Comparing to other resistive spin Hall metal (W, Ta, *etc.*), Pt is a particularly attractive candidate for SOT device applications due to its relatively low resistivity and large intrinsic spin Hall conductivity from its band structure[32]. However, the reported $\xi_{DL}^{\mathrm{Pt}} = 0.07{\sim}0.12$ is still quite low in several Pt/FM systems[14, 42], and the price concern from scarcity also makes it less attractive in mass production than Ta or W; the dopant Cu, rather than other resistive impurities used in previous works, is conductive, inexpensive, and widely-adopted in modern CMOS process, and thus can serve as an ideal scattering impurity. Despite several works have shown that Pt-Cu alloy have larger spin Hall angle than Pt[43, 44], a direct demonstration on its SOT switching performance has not yet been investigated. We first ensure PMA can be obtained in $Pt_xCu_{1-x}$(5)/Co(1)/MgO(2) heterostructures, then determine the DL-SOT efficiency $\xi_{DL}^{\mathrm{Pt-Cu}}$ by hysteresis loop shift measurements. Within the PMA regime, tunable $\xi_{DL}^{\mathrm{Pt-Cu}}$ and coercivity are observed. The enhanced $\xi_{DL}^{\mathrm{Pt-Cu}}$ resulting from the raised $\rho_{xx}^{\mathrm{Pt-Cu}}$ by optimal alloying of Pt is ~ 0.44 for $Pt_{0.57}Cu_{0.43}$, and the obtained $\sigma_{SH}^{\mathrm{Pt-Cu}}$ is about $5.38\times10^5$ $(\hbar/2e)$ $\Omega^{-1}\cdot m^{-1}$. We furthermore perform current-induced magnetization switching on heterostructures with PMA. Deterministic SOT switching is achieved, and the lowest critical switching current density in $Pt_xCu_{1-x}$ layer is ~ $2.37\times10^6$ $A\cdot cm^{-2}$ for $Pt_{0.57}Cu_{0.43}$. This large reduction on critical switching current is attributed to the concurrent enhancement of $\xi_{DL}$ and the reduction of coercivity by alloying. Also, good thermal stability is maintained within the PMA regime, and the lowest zero-thermal critical switching current is ~ $1.12\times10^7$ $A\cdot cm^{-2}$ for $Pt_{0.57}Cu_{0.43}$. The lowest power consumption without current shunting is ~ $4.64\times10^{11}$ $mW\cdot cm^{-3}$ for $Pt_{0.57}Cu_{0.43}$, significantly lower than that from a pure Pt control sample ($1.61\times10^{13}$ $mW\cdot cm^{-3}$). We further provide a comprehensive benchmarking summary of SOT switching power consumption in various materials systems, in which we show that $Pt_{0.57}Cu_{0.43}$ is a tentative spin current source with low power consumption, moderate resistivity, and minimal shunting effect.

## 2. Materials preparation and characterization

### 2.1. sample structures, preparation methods and characterization

As illustrated in **Figure 1a**, $Pt_xCu_{1-x}$(5)/Co(1)/MgO(2) heterostructure are sputter-deposited onto $SiO_2$ substrate with Ar flow of 3mTorr at room temperature. The numbers in parenthesis represent the layer thickness in nanometers, and x is the atomic percentage Pt concentration. All the samples are capped by Ta(2), which is expected to be fully oxidized under ambient atmosphere and therefore its SOT contribution is negligible. The composition of $Pt_xCu_{1-x}$ is controlled by co-sputtering pure Pt and pure Cu sources under calibrated sputtering power. The saturation magnetization of Co layer is characterized by a vibrating sample magnetometer (VSM) to be 1414 emu·cm$^{-3}$. For the purpose of electrical characterization, the samples are lithographically patterned into Hall bar devices (5 μm × 60 μm). The average longitudinal resistivity ($\rho_{xx}$) of $Pt_xCu_{1-x}$ varying with Pt concentration are then characterized via typical four-point probe resistance measurements on patterned $Pt_xCu_{1-x}$(5) devices, as shown in **Figure 1b**. The $\rho_{xx}$ of $Pt_xCu_{1-x}$ increases with more Cu concentration in Pt-Cu alloy from 27.4 μΩ·cm (pure Pt) to 122.5 μΩ·cm ($Pt_{0.39}Cu_{0.61}$), then decreases to 32.7 μΩ·cm (pure Cu), showing a typical resistivity trend of well-mixed alloy that resistivity is proportional to alloy randomness. The resistivity of Co is characterized to be 26.6 μΩ·cm.

### 2.2. PMA window

The magnetic anisotropy properties of $Pt_xCu_{1-x}$(5)/Co(1)/MgO(2) heterostructure are characterized by anomalous Hall effect (AHE) hysteresis loops, as illustrated in **Figure 1a**. By reading Hall resistance ($R_H$) under varying out-of-plane external field ($H_z$) with a fixed in-plane bias current ($I$ = 1.9 mA), the perpendicular/in-plane magnetization anisotropy (PMA/IMA) degree can be defined as the ratio between remnant magnetization ($M_R$) and saturated magnetization ($M_S$), which are evaluated by $R_H$ under zero $H_z$ and maximum $R_H$ respectively; and the coercive field ($H_c$) is defined as the field when normalized $R_H$ changes its sign. As shown in **Figure 1c-d**, the PMA/IMA window respectively locate in the Pt/Cu-rich regime, and the lowest threshold Pt concentration to maintain perfect PMA is about 50%. At the same time, the coercivity of Co layer reduces with increasing Cu content in the Pt-Cu alloy. Both features are consistent with the trend that alloying by Cu would weaken the overall SOC of Pt-Cu alloy.

## 3. Experimental results

### 3.1. damping-like torque efficiency of Pt-Cu alloy

We first determine the DL-SOT efficiency ($\xi_{DL}$) of Pt-Cu alloy by current-induced hysteresis loop-shift measurements [45] for $Pt_xCu_{1-x}$(5)/Co(1)/MgO(2) heterostructures with PMA, which is illustrated in **Figure 2a**. Based on a DL-SOT + homochiral Néel domain wall motion scenario[46], the magnetization experiences a perpendicular effective field ($H_{eff}^{z}$) induced by in-plane charge current ($I_{applied} \parallel \hat{x}$), which results in a shift of the out-of-plane hysteresis loop. $\xi_{DL}$ can be estimated by $H_{eff}^{z}/I$ from linear fits to $H_{eff}^{z}$ versus $I_{applied}$. Under this scenario, $\xi_{DL}$ reaches a saturated value when the external in-plane field $H_x$ just overcome the Dzyaloshinskii-Moriya interaction (DMI) effective field $|H_{DMI}|$ originated from the $Pt_xCu_{1-x}$/Co interface and then realigns the chiral domain wall moments, as shown in **Figure 2b-d**. Moreover, $\xi_{DL}$ derived from a macrospin model can be written as[45, 47]

$$\xi_{DL} = \left(\frac{2}{\pi}\right)\frac{2e}{\hbar}\mu_0 M_s t_{\mathrm{Co}} w t_{\mathrm{Pt-Cu}}(1+s)\left(\frac{H_{eff}^{z}}{I}\right), \tag{2}$$

where $\mu_0$ is vacuum permeability, $M_s$ and $t_{\mathrm{Co}}$ are the magnetization and thickness of the Co layer respectively. $s \equiv I_{\mathrm{FM}}/I_{\mathrm{NM}} = t_{\mathrm{Co}}\rho_{\mathrm{Pt-Cu}}/t_{\mathrm{Pt-Cu}}\rho_{\mathrm{Co}}$ is the current shunting parameter; *w* and *d* are the width and thickness of $Pt_xCu_{1-x}$ layer respectively. Based on the $H_{eff}^{z}/I$ results obtained from hysteresis loop-shift measurements, the relationship between $\xi_{DL}$ and $\rho_{\mathrm{Pt-Cu}}$ is summarized in **Figure 2e**. The largest $\xi_{DL}^{\mathrm{Pt-Cu}} \sim 0.44$ for $Pt_{0.57}Cu_{0.43}$ (with 82.5 μΩ·cm) is more than two times larger than $\xi_{DL}^{\mathrm{Pt}}$~0.14 for pure Pt (with 27.4 μΩ·cm). Besides the doped HM itself, the interfacial condition such as interface roughness[48] and spin transparency[49] might also affect the performance. Despite the interface quality might be altered during alloying, $\xi_{DL}$ varies fairly linearly with respect to $\rho_{\mathrm{Pt-Cu}}$ within the PMA regime, and the lower-bounded spin Hall conductivity ($\sigma_{SH}^{\mathrm{Pt-Cu}}$) is estimated to be 5.38×10$^5$($\hbar/2e$) Ω$^{-1}$·m$^{-1}$, close to that of pure Pt. This linear relationship indicates the bulk intrinsic Pt property is preserved during alloying, and the influence from the interface plays a minor role in this case. It is interesting to note that, the preservation on intrinsic Pt property of Pt-Cu is quite different from the case of Pt-Hf, where $\sigma_{SH}^{\mathrm{Pt-Hf}}$ is reduced as the Hf concentration goes beyond 12.5%[35].

### 3.2. current-induced magnetization switching

We then perform current-induced magnetization switching measurements on the $Pt_xCu_{1-x}$(5)/Co(1)/MgO(2) heterostructures with PMA. The measurement setup is illustrated in **Figure 3a**. We alternatively apply in-plane longitudinal write/read current pulses $I_{write/read}$ to control/sense the magnetization by DL-SOT/AHE, with the aid of an external in-plane field along $\hat{x}$ to overcome the DMI effective field. The pulse width ($t_{pusle}$) of $I_{write}$ is set as 0.05 s; $I_{read}$ is set as 0.03 mA. The switching current ($I_{sw}$) is defined as the write current at which the normalized Hall resistance ($R_H$) changes sign; and the saturated external field ($H_{sat}$) is defined as the applied in-plane field at which the switching current reaches its minimum ($I_c = I_{sw}^{min}$). As the results shown in **Figure 3b-d**, deterministic current-induced magnetization switching is demonstrated, and the reversing switching polarity with opposite external field direction is consistent with the SHE+DMI scenario.

For the control experiments (pure Pt-based device), $I_c$ is about 7.30 mA and $J_c$ in Pt layer is about 2.42×10$^7$ A·cm$^{-2}$, which are close to previously reported results[3]. And for Pt-Cu alloy, the lowest $I_c$ is about 0.96 mA and $J_c$ in the $Pt_{0.57}Cu_{0.43}$ layer is about 2.37×10$^6$ A·cm$^{-2}$ for the $Pt_{0.57}Cu_{0.43}$-based device. This low current density is comparable to other reported conventional and emergent spin current source materials at room temperature, such as β-Ta(4)/CoFeB(1) (5.47 ×10$^6$ A·cm$^{-2}$)[1] 、BiSb(5)/MnGa(3) (1.10×10$^6$ A·cm$^{-2}$)[22]. The trend of $I_c$ and $H_{sat}$ as changing Pt concentration is as expected and consistent with the observed trend of $\xi_{DL}$ and $H_{DMI}$ from loop-shift measurements: the larger $\xi_{DL}$, the lower $I_c$, and the smaller $H_{DMI}$, the smaller $H_{sat}$. Also note that the coercive field is reduced when the Pt content is reduced. Therefore, we attribute this significant reduction of $J_c$ to the simultaneous enhancement of spin-orbit torque efficiency and reduction of coercivity of the Co layer by proper alloying.

### 3.3. thermal stability and power consumption

The effect of alloying on the thermal stability of heterostructures-of-interest has been largely ignored in previous studies. Since DL-SOT switching is a thermally-activated process with the write current pulse we applied, the thermal stability of the Co layer can be determined by performing switching measurements with different pulse widths. Based on a thermal-assisted model, the relationship between $I_c$ and $t_{pusle}$ follows[50]

$$I_c = I_{c0}\left[1 - \frac{1}{\Delta}\ln\left(\frac{t_{pulse}}{\tau_0}\right)\right], \tag{3}$$

where $I_{c0}$ is the critical switching current in the absence of Joule heating; $\Delta = U/k_B T$ is thermal stability factor representing the energy barrier between UP and DOWN state of the FM layer with PMA; $t_{pulse}$ is the write pulse width which ranges from 0.05 s to 1 s; $\tau_0 \sim 1$ ns is the intrinsic thermal attempt time. By varying $t_{pulse}$, the $I_{c0}$ and $\Delta$ with different Pt concentration can be extracted and estimated from the intercept and slope by linear fits of $I_c$ versus $\ln(t_{pulse}/\tau_0)$, as summarized in **Figure 4a-b**. We find the lowest $|I_{c0}|$~$3.09 \pm 0.08$ mA for $Pt_{0.57}Cu_{0.43}$, and $|I_{c0}|$~$11.88 \pm 1.10$ mA for pure Pt in our control experiments. Moreover, the trend of $I_{c0}$ is consistent with the DL-SOT efficiency of Pt-Cu alloy, which reconfirms its SHE origin. It is noteworthy that no serious thermal stability degradation is observed despite alloying, with $\Delta = 28.85 \pm 1.15$ for $Pt_{0.57}Cu_{0.43}$ and $\Delta = 32.72 \pm 1.27$ for pure Pt. This thermal tolerance against alloying is beneficial for data retention, and ensure the performance of the SOT device.

Next, we calculate the upper bound of write power consumption per single bit per volume without current shunting and Joule heating ($p_0 = \rho J_{c0}^2$) and switching efficiency ($\varepsilon \equiv \Delta / I_{c0}$), where $\rho$ and $J_{c0}$ are the resistivity and critical zero-thermal switching current density in the spin current source channel; $\Delta$ and $I_{c0}$ are the thermal stability and zero-thermal critical switching current of the whole device. As summarized in **Figure 4c**, the lowest $p_0$ is about $4.81 \times 10^{12}$ mW·cm$^{-3}$ for $Pt_{0.57}Cu_{0.43}$, which is reduced by almost an order from that for pure Pt ($4.25 \times 10^{13}$ mW·cm$^{-3}$). This reduction on power consumption is the result of lower $J_{c0}$ and moderate resistivity of the Pt-Cu alloy. If the thermal effect is further considered, the apparent power consumption ($p = \rho J_c^2$) will become even lower due to the reduced $J_c$ ($4.64 \times 10^{11}$ mW·cm$^{-3}$ for $Pt_{0.57}Cu_{0.43}$ vs. $1.60 \times 10^{13}$ mW·cm$^{-3}$ for pure Pt). In addition, the largest $\varepsilon = 9.33 \pm 0.14$ mA$^{-1}$ for $Pt_{0.57}Cu_{0.43}$ is more than three times better than $\varepsilon = 2.75 \pm 0.15$ mA$^{-1}$ for pure Pt, which is due to the preserved $\Delta$ and lower $I_{c0}$.

## 4. Benchmarking performance among Pt alloys

Finally, we also compare the SOT switching power consumption using Pt-Cu alloy with other common materials systems, including Pt-based[33, 34, 36-38], β-Ta[1], β-W[51] and chalcogenide-based[19, 20, 22] spin current sources (SCSs). Note that we only focus on works with room-temperature switching results. Given that rare works studied on $J_{c0}$, we estimate power consumption by the apparent $p = \rho J_c^2$. If we further consider the energy dissipation factor (η, unitless) due to current shunting effect, the actual power consumption should be proportional to $(1 + \eta)$ with

$$1 + \eta = (1 + s)\frac{t_{\mathrm{SCS}}}{t_{\mathrm{FM}} + t_{\mathrm{SCS}}}, \tag{4}$$

where $s \equiv I_{\mathrm{FM}}/I_{\mathrm{SCS}} = t_{\mathrm{FM}}\rho_{\mathrm{SCS}}/t_{\mathrm{SCS}}\rho_{\mathrm{FM}}$ is the shunting parameter, $t$ and $\rho$ are thickness and resistivity, respectively (see supporting information for detailed derivations). Replacing FM to magnetic insulator is viable to achieve negligible energy dissipation from avoiding current shunting through FM layer[52, 53]. But here we estimate η based on CoFeB, a much more common FM for industrial purpose, with a specific heterostructure (SCS(5)/CoFeB(1) with $\rho_{\mathrm{CoFeB}} = 190\ \mu\Omega \cdot \mathrm{cm}$). The calculated power consumption and η are summarized in **Figure 5**. It is noteworthy that despite low power consumption can be achieved due to the large DL-SOT efficiency of chalcogenide-based SCSs, the enhanced energy dissipation due to current shunting from the large resistivity of SCSs actually would make it energetically unfavorable. On the other hand, despite Pt-based SCSs would not suffer from serious energy dissipation due to relatively low resistivity, the raised resistivities accompanying with alloying or thin layer insertion might compensate the merits from lower critical switching currents, even make it worse in several cases. However, $Pt_{0.57}Cu_{0.43}$ in this work not only significantly improve the power consumption of pure Pt (even the best among reported Pt-based SCSs), but also is competitive to other SCSs. With low power consumption and moderate resistivity, $Pt_{0.57}Cu_{0.43}$ therefore is competitive candidate as SCS in future SOT-MRAM.

## 5. Conclusion

To summarize, we systematically investigate on $Pt_xCu_{1-x}$/Co/MgO heterostructure with PMA. Large enhancement on DL-SOT efficiency of Pt-Cu alloy is demonstrated by hysteresis loop shift measurements, resulting from tuning resistivity by proper alloying on intrinsic Pt. The largest DL-SOT efficiency is up to 0.44 for $Pt_{0.57}Cu_{0.43}$, when the resistivity is raised from 27.4 μΩ·cm to 82.5 μΩ·cm. And the lower-bound of the spin Hall conductivity of Pt-Cu alloy is about $5.38\times 10^5(\hbar/2e)\ \Omega^{-1}\cdot m^{-1}$, similar to that of intrinsic Pt. This large DL-SOT efficiency is then confirmed by current-induced magnetization switching measurements, and contributes to an ultra-low critical switching current density together with coercivity reduction due to alloying. The lowest critical switching current density is about $2.37\times10^6\ A\cdot cm^{-2}$ in the $Pt_{0.57}Cu_{0.43}$ layer. Furthermore, thermal stability of the Co layer is unaffected by alloying. Finally, Pt-Cu alloy can significantly reduce the power consumption due to a large DL-SOT efficiency and moderate resistivity. The lowest power consumption excluding current shunting is about $4.64\times10^{11}\ mW/cm^3$ for $Pt_{0.57}Cu_{0.43}$ ($1.61\times10^{13}\ mW\cdot cm^{-3}$ for pure Pt). This low power is the best performance among various reported Pt-based spin current sources, and competitive to other emergent spin current sources. Moreover, the moderate resistivity of Pt-Cu alloy can mitigate energy dissipation due to current shunting. Therefore, Pt-Cu alloy is expected to be an attractive candidate as the spin current source in future SOT-MRAM applications.

## 6. Experimental section

### 6.1. sample growth and characterizations

The $Pt_xCu_{1-x}$(5)/Co(1)/MgO(2) heterostructures with different Pt concentrations (x) were sputter-deposited onto $SiO_2$ substrates in a customized ultra-high vacuum magnetron sputtering system with base pressure of $5\times10^{-8}$ Torr. The films were deposited via DC or RF magnetron sputtering at room temperature and in an Ar growth pressure of 3 mTorr. All the films were capped by Ta(2) as protective layer, which was expected to be fully oxidized under ambient atmosphere. The saturation magnetization of Co was characterized by a vibrating sample magnetometer (VSM).

### 6.2. device fabrication and measurements

All the heterostructures were fabricated into micro-size Hall bar (5 μm × 60 μm) by standard photolithography for electrical measurements. The resistivities were characterized by standard four-point probe measurements. Anomalous Hall resistances were measured by a home-made probe station, which is capable to simultaneously apply in-plane and out-of-plane magnetic field. The electrical measurements were performed with a dc current source (by Keithley 2400) and a voltage meter (by Keithley 2000).

## Supporting Information

Details of power consumptions from various references can be found in Supporting Information.

## Acknowledgements

This work is supported by the Ministry of Science and Technology of Taiwan (MOST) under grant No. MOST-108-2636-M-002-010 and by the Center of Atomic Initiative for New Materials (AI-Mat), National Taiwan University from the Featured Areas Research Center Program within the framework of the Higher Education Sprout Project by the Ministry of Education (MOE) in Taiwan under grant No. NTU-108L9008. We would also like to thank Dr. Lijun Zhu for fruitful discussions on the estimation of SOT power consumption in various materials systems.

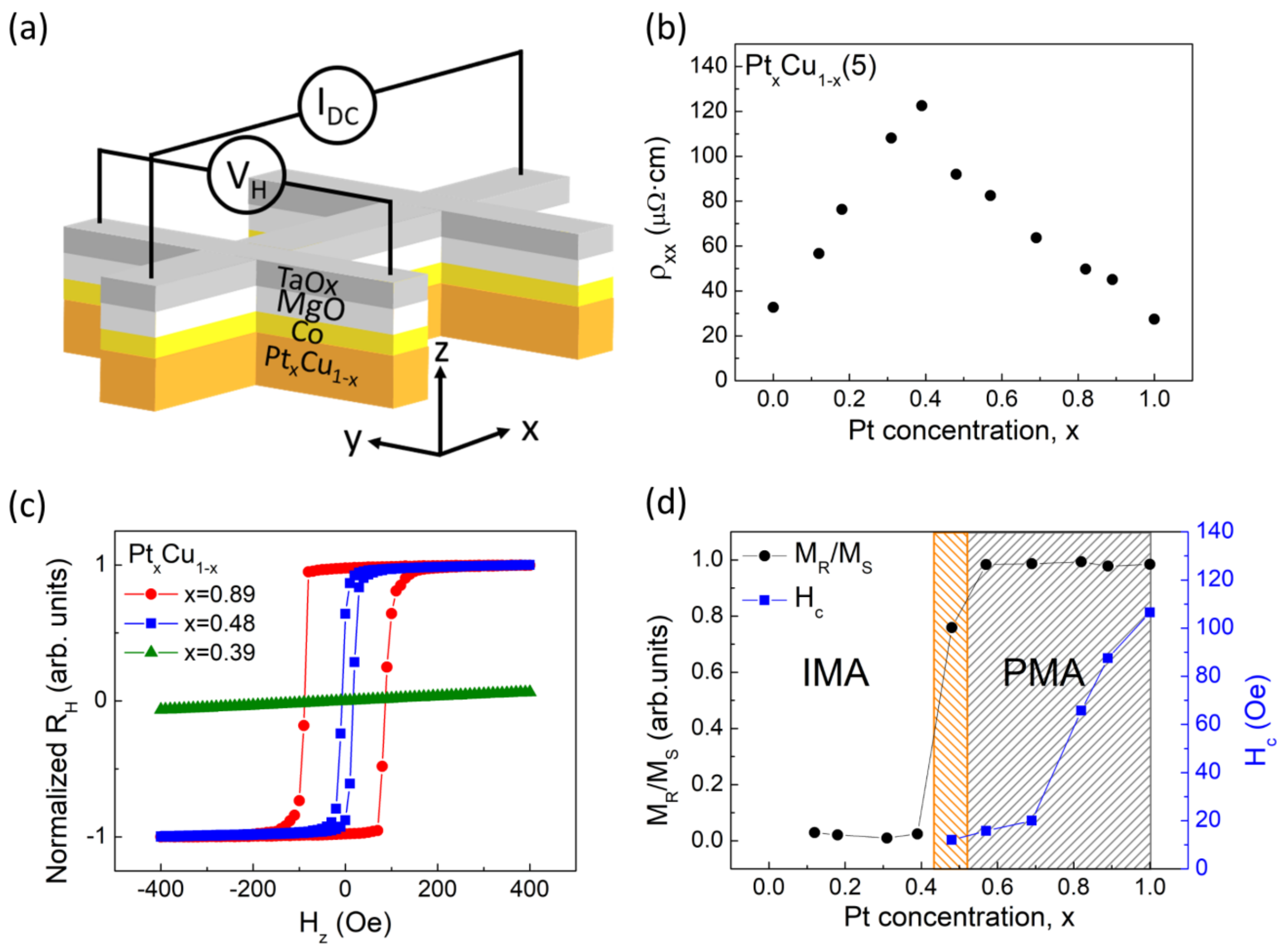

**Figure 1. Experimental setup and magnetic properties of $Pt_xCu_{1-x}(5)/Co(1)/MgO(2)$ heterostructures. a,** schematics of patterned heterostructures and circuits setup. **b,** resistivity of $Pt_xCu_{1-x}(5)$ with varying Pt concentration (denoted as x). **c,** representative anomalous Hall resistance hysteresis loops for $Pt_xCu_{1-x}(5)/Co(1)/MgO(2)$ heterostructures with x = 0.89, 0.48, and 0.39. **d,** out-of-plane remnant/saturated magnetization and coercivity of $Pt_xCu_{1-x}(5)/Co(1)/MgO(2)$ heterostructures as functions of Pt concentration.

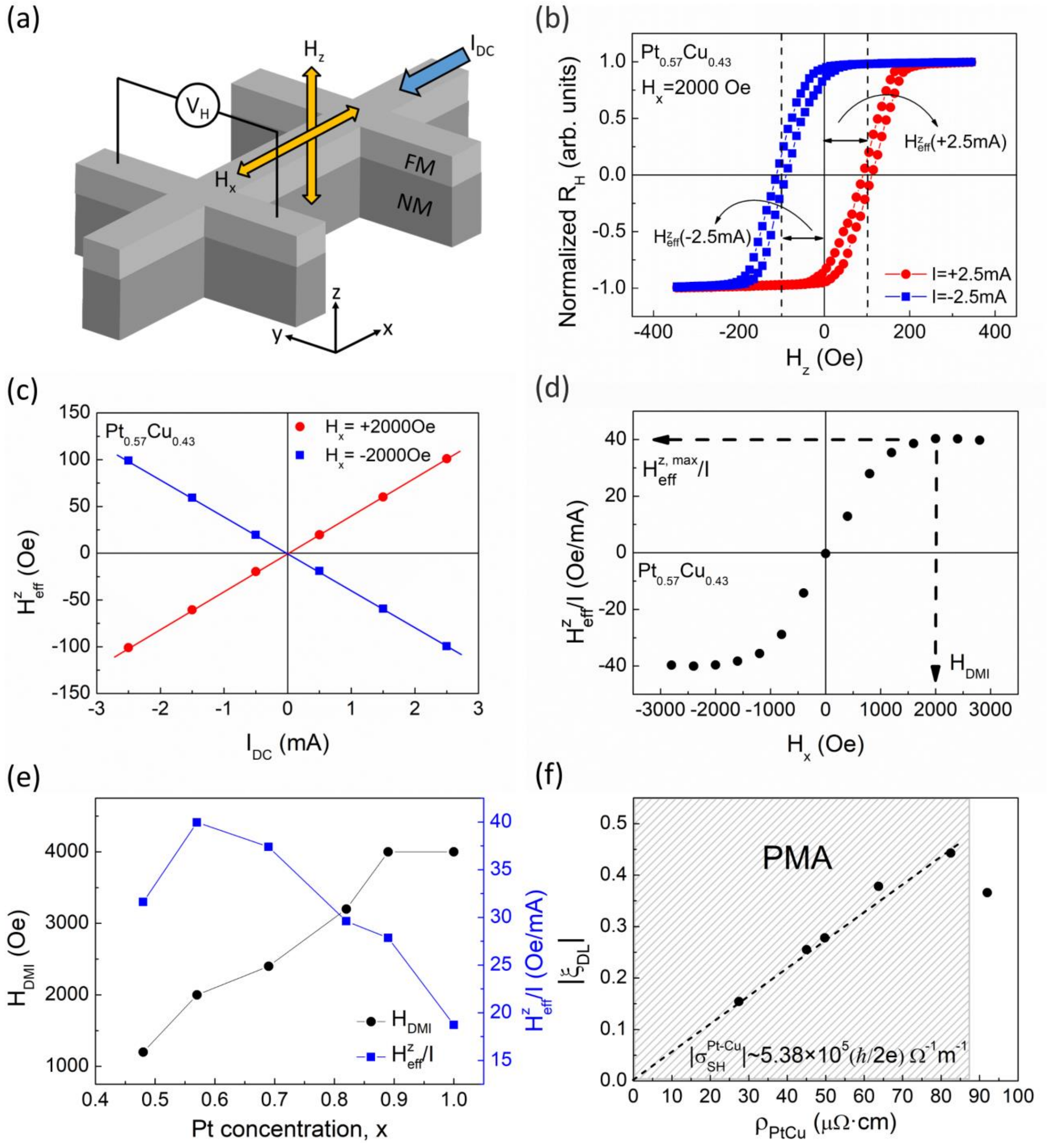

**Figure 2. Damping-like torque efficiency characterization. a,** schematics of a Hall bar device for anomalous Hall hysteresis loop shift measurements, where $I_{DC}$ represents the in-plane current along $\hat{x}$ . $H_x$ and $H_z$ are the applied in-plane/out-of-plane magnetic field, respectively. **b,** representative hysteresis loop shift results of a $Pt_{0.57}Cu_{0.43}$-based device under $H_x = 2000$ Oe and $I_{DC} = \pm 2.5$ mA. **c,** $H_{eff}^{z}$ as functions of $I_{DC}$ of a $Pt_{0.57}Cu_{0.43}$-based device under $H_x = \pm 2000$ Oe. **d,** $H_{eff}^{z}/I$ as a function of $H_x$, where $H_{eff}^{z,max}/I$, $H_{DMI}$ are the saturated $H_{eff}^{z}/I$ and the corresponding $H_x$. **e,** the summary of $H_{eff}^{z}/I$ and $H_{DMI}$ versus Pt concentration. **f,** calculated damping-like torque efficiency ($\xi_{DL}$) versus longitudinal resistivity ($\rho_{xx}$) of Pt-Cu alloy, where the slashed area corresponds to samples with PMA.

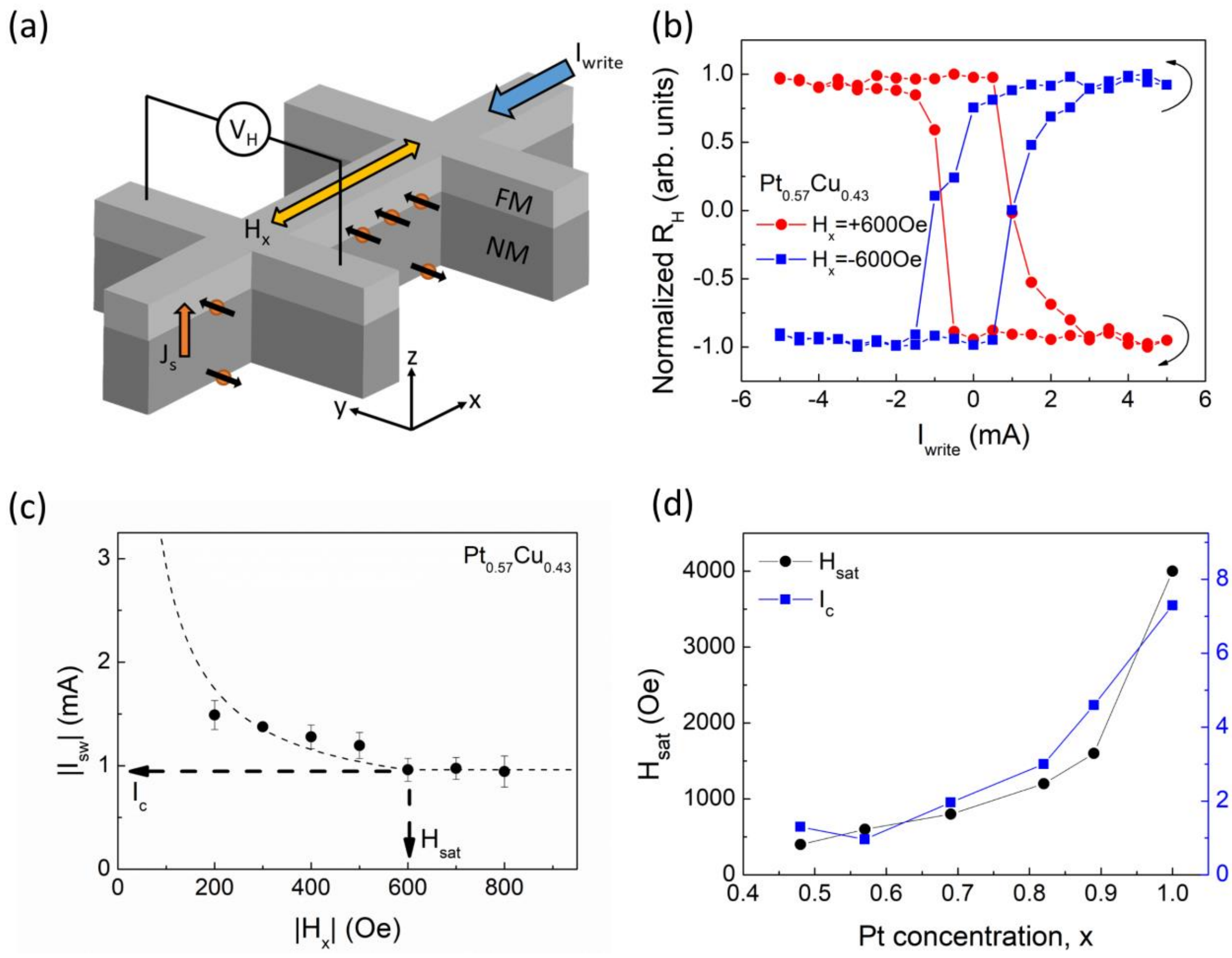

**Figure 3. Current-induced magnetization switching. a,** schematic illustration of current-induced magnetization switching measurements, where $I_{write}$ is the write current along $\hat{x}$, $H_x$ is the in-plane external field along $\hat{x}$, $J_s$ is the spin current with spin polarization along $\hat{y}$. **b,** representative current-induced magnetization loops of a $Pt_{0.57}Cu_{0.43}$-based device under $H_x = \pm 600$ Oe . **c,** switching current ($|I_{sw}|$) versus $|H_x|$, where the critical switching current ($I_c$) and saturation field ($H_{sat}$) are the saturated $|I_{sw}|$ and its corresponding $|H_x|$. **d,** the summary of $I_c$ and $H_{sat}$ with varying Pt concentration.

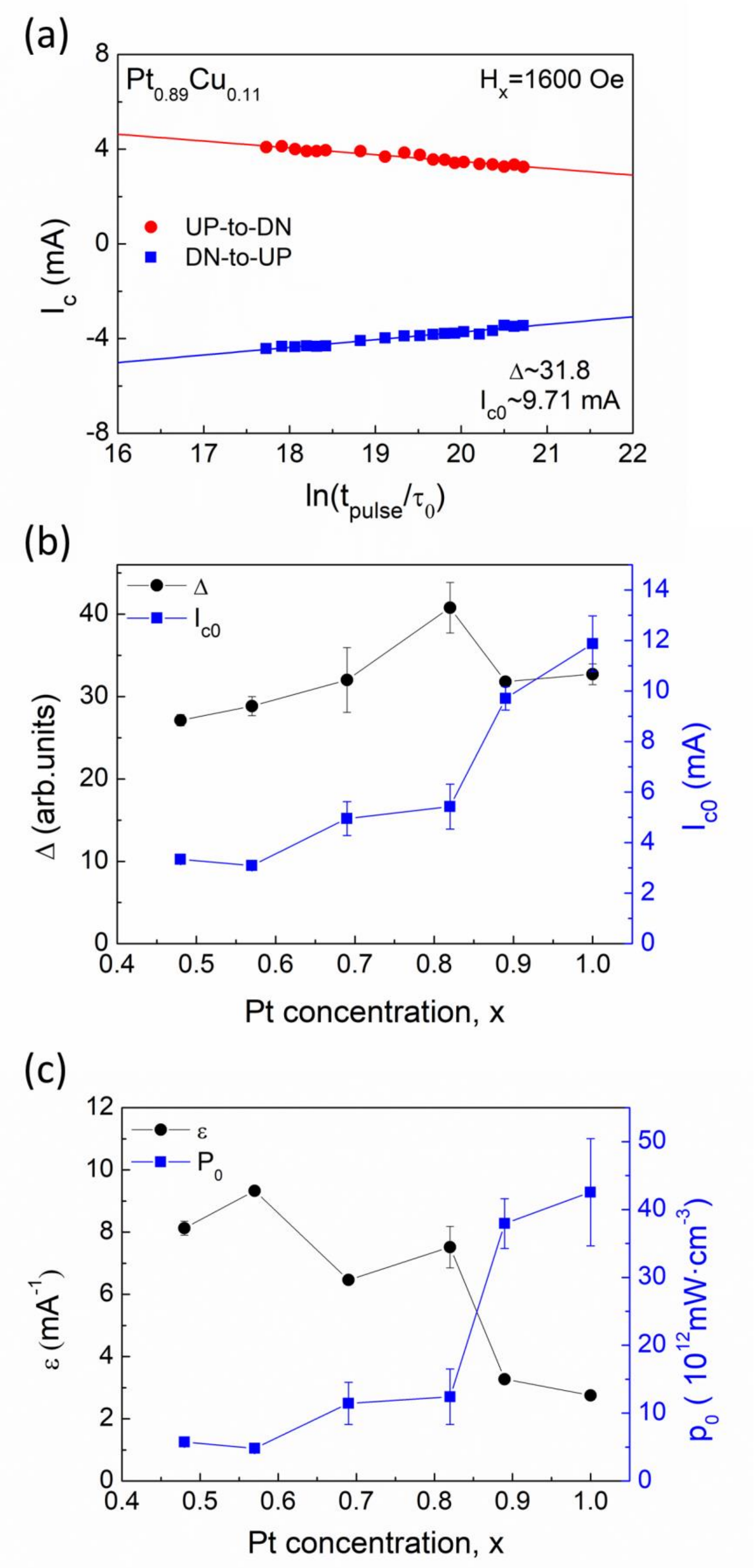

**Figure 4. Thermal stability and power consumption. a,** critical switching current ($I_c$) versus pulse width ($t_{pulse}$) of a representative $Pt_{0.57}Cu_{0.43}$-based device under $H_x = 1600$ Oe. **b,** summary of thermal stability ($\Delta$) of the Co layer and zero-thermal critical switching current ($I_{c0}$) with different Pt concentration. **c,** summary of switching efficiency ($\varepsilon$) and power consumption without Joule heating ($p_0$) with different Pt concentration.

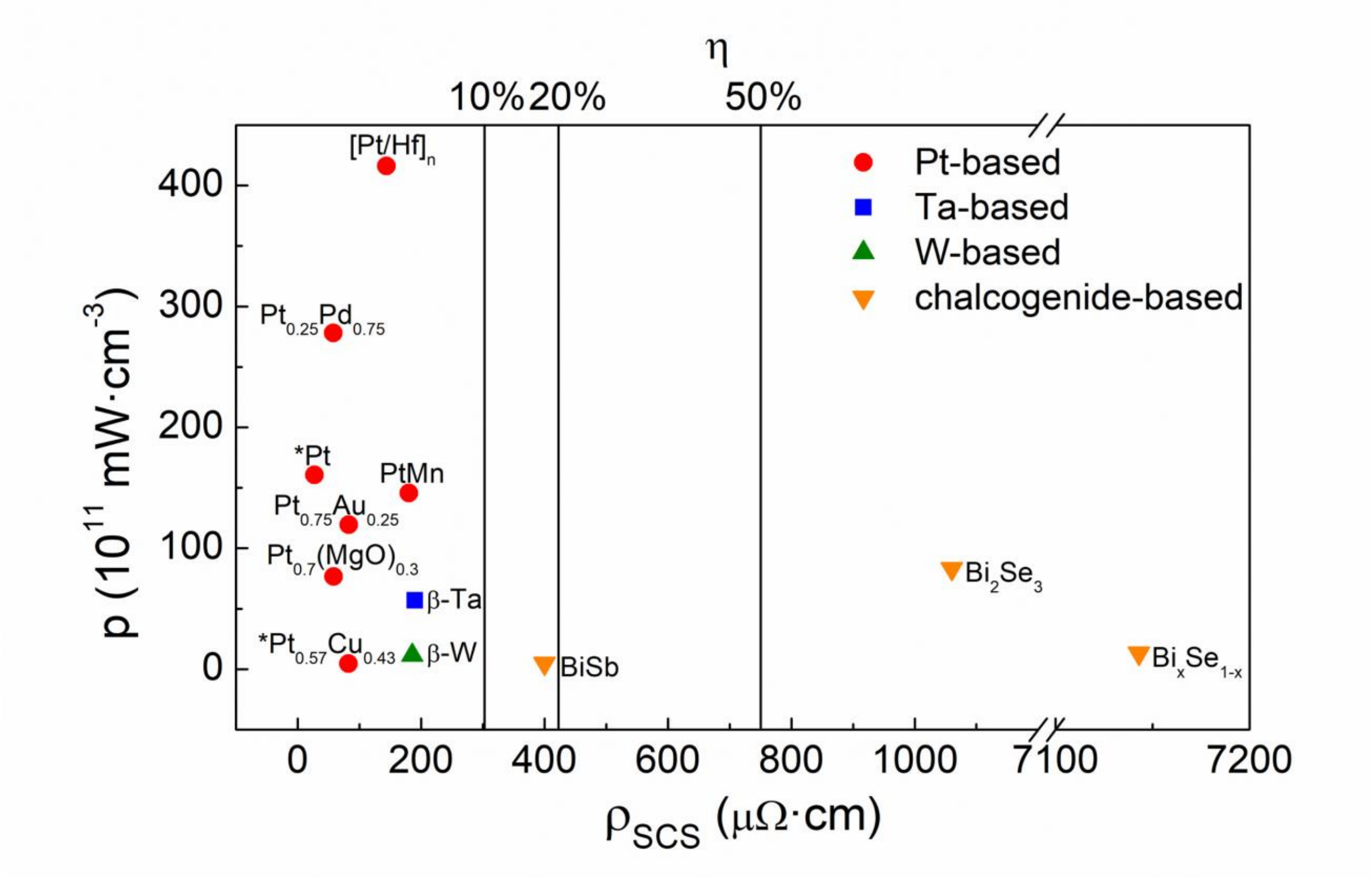

**Figure 5. Benchmarking of switching power consumption.** The summary of power consumption ($p$) for room-temperature current-induced SOT switching versus longitudinal resistivity for various spin current sources ($\rho_{\mathrm{SCS}}$). The starred points represent results from this work. The upper x-axis indicates the energy dissipation proportion factor (η) based on a SCS(5)/CoFeB(1) geometry.

# Supporting Information

**Benchmarking of spin-orbit torque switching efficiency in Pt alloys**

*Chen-Yu Hu, Chi-Feng Pai**

## 1. Power consumption of a SOT-switched ferromagnet

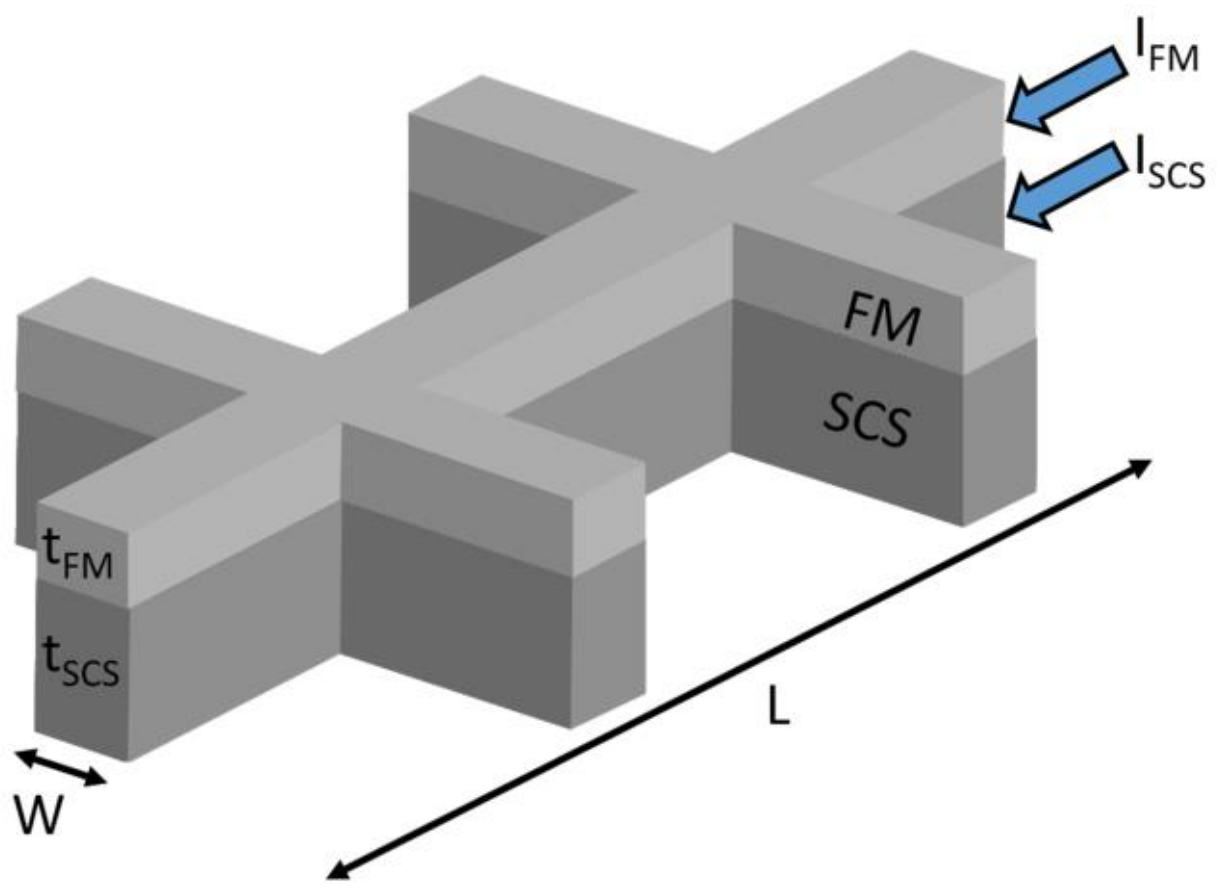

**Figure S1.** Schematic illustration of SOT device geometry and parameters definition.

For a simple double layer structure including a spin current source (SCS) channel with an adjacent ferromagnetic (FM) layer, the critical switching current of whole device ($I_c$) can be expressed as

$$I_c = (1+s)I_c^{\mathrm{SCS}} = (1+s)J_c^{\mathrm{SCS}} W t_{\mathrm{SCS}}, \qquad \text{(S1)}$$

where $s$ is the shunting parameter and is defined as $s \equiv I_{\mathrm{FM}}/I_{\mathrm{SCS}} = t_{\mathrm{FM}}\rho_{\mathrm{SCS}}/t_{\mathrm{SCS}}\rho_{\mathrm{FM}}$; $I_c^{\mathrm{SCS}}$ and $J_c^{\mathrm{SCS}}$ are the charge current and charge current density flowing in the SCS channel, respectively; $W$ is the device width; $t_{\mathrm{SCS}}$ is the SCS thickness; $\rho_{\mathrm{SCS}}$ and $\rho_{\mathrm{FM}}$ respectively are the resistivities of SCS and FM. Moreover, based on parallel resistance assumption, the effective resistance of whole device ($R$) can be written as

$$R = (R_{\mathrm{FM}}^{-1} + R_{\mathrm{NM}}^{-1})^{-1} = \left(\frac{t_{\mathrm{FM}}W}{\rho_{\mathrm{FM}}L} + \frac{t_{\mathrm{SCS}}W}{\rho_{\mathrm{NM}}L}\right)^{-1} = \frac{L}{W}\left(\frac{t_{\mathrm{FM}}}{\rho_{\mathrm{FM}}} + \frac{t_{\mathrm{SCS}}}{\rho_{\mathrm{NM}}}\right)^{-1} = \frac{L}{W}\frac{\rho_{\mathrm{SCS}}}{t_{\mathrm{SCS}}}\frac{1}{1+s}, \qquad \text{(S2)}$$

where $R_{\mathrm{FM}}$ and $R_{\mathrm{SCS}}$ are the resistances of FM layer and SCS channel, respectively; $L$ is the device length. Thus, the critical write power consumption per single bit per volume ($p_{write}$) can be expressed as

$$p_{write} = \rho {J_c}^2 = \frac{{I_c}^2 R}{WL(t+d)} = (1+s)\frac{t_{\mathrm{SCS}}}{t_{\mathrm{FM}} + t_{\mathrm{SCS}}}\rho_{\mathrm{SCS}} {J_c^{\mathrm{SCS}}}^2, \qquad \text{(S3)}$$

where $\rho$ and $J_c$ are the effective resistivity and critical switching current density of the whole device. To systematically compare power consumption among different materials systems by using published data and the data taken in this work, we exclude the contribution of current shunting effect and use $p = \rho_{\mathrm{SCS}} {J_c^{\mathrm{SCS}}}^2$ as the measure of power consumption. The power extracted from several representative works on room-temperature SOT switching are summarized in **Table S1**, including various Pt-based, Ta-based, W-based, and chalcogenide-based SCSs.

| SCS type | Materials systems | $\rho_{\mathrm{SCS}}$ [μΩ·cm] | $J_c^{\mathrm{SCS}}$ [A·cm$^{-2}$] | $p = \rho_{\mathrm{SCS}} {J_c^{\mathrm{SCS}}}^2$ [mW·cm$^{-3}$] | Reference |
|---|---|---|---|---|---|
| Pt-based | $Pt_{0.57}Cu_{0.43}$/Co/MgO | 82.5 | $2.37 \times 10^6$ | $4.64 \times 10^{11}$ | This work |
| Pt-based | Pt/Co/MgO | 27.4 | $2.42 \times 10^7$ | $1.61 \times 10^{13}$ | This work |
| Pt-based | PtMn/Hf/CoFeB/MgO | 180 | $9 \times 10^6$ | $1.46 \times 10^{13}$ | [S1] |
| Pt-based | $Au_{0.25}Pt_{0.75}$/Co/MgO | 83 | $1.2 \times 10^7$ | $1.20 \times 10^{13}$ | [S2] |
| Pt-based | $Pt_{0.25}Pd_{0.75}$/Co | 57.5 | $2.2 \times 10^7$ | $2.78 \times 10^{13}$ | [S3] |
| Pt-based | $Pt_{0.7}(MgO)_{0.3}$/Co | 58 | $1.15 \times 10^7$ | $7.67 \times 10^{12}$ | [S4] |
| Pt-based | $[Pt/Hf]_n$/Co/MgO | 144 | $1.7 \times 10^7$ | $4.16 \times 10^{13}$ | [S5] |
| Ta-based | β-Ta/CoFeB/MgO | 190 | $5.47 \times 10^6$ | $5.69 \times 10^{12}$ | [S6] |
| W-based | W/CoFeB | 185.7 | $2.5 \times 10^6$ | $1.16 \times 10^{12}$ | [S7] |
| chalco.-based | $Bi_2Se_3$/CoTb | 1060 | $2.8 \times 10^6$ | $8.31 \times 10^{12}$ | [S8] |
| chalco.-based | MgO/$Bi_xSe_{1-x}$/Ta/CoFeB/Gd/CoFeB/MgO | 7143 | $4.3 \times 10^5$ | $1.32 \times 10^{12}$ | [S9] |
| chalco.-based | BiSb/MnGa | 400 | $1.1 \times 10^6$ | $4.84 \times 10^{11}$ | [S10] |

**Table S1.** Summary of $\rho_{\mathrm{SCS}}$, $J_c^{\mathrm{SCS}}$ and calculated $p = \rho_{\mathrm{SCS}} {J_c^{\mathrm{SCS}}}^2$ of reported spin current sources in different materials systems. All these materials systems are with perpendicular magnetic anisotropy (PMA) and the switching measurements were performed at room temperature.

## 2. Power consumption estimation of an ideal SOT-controlled device

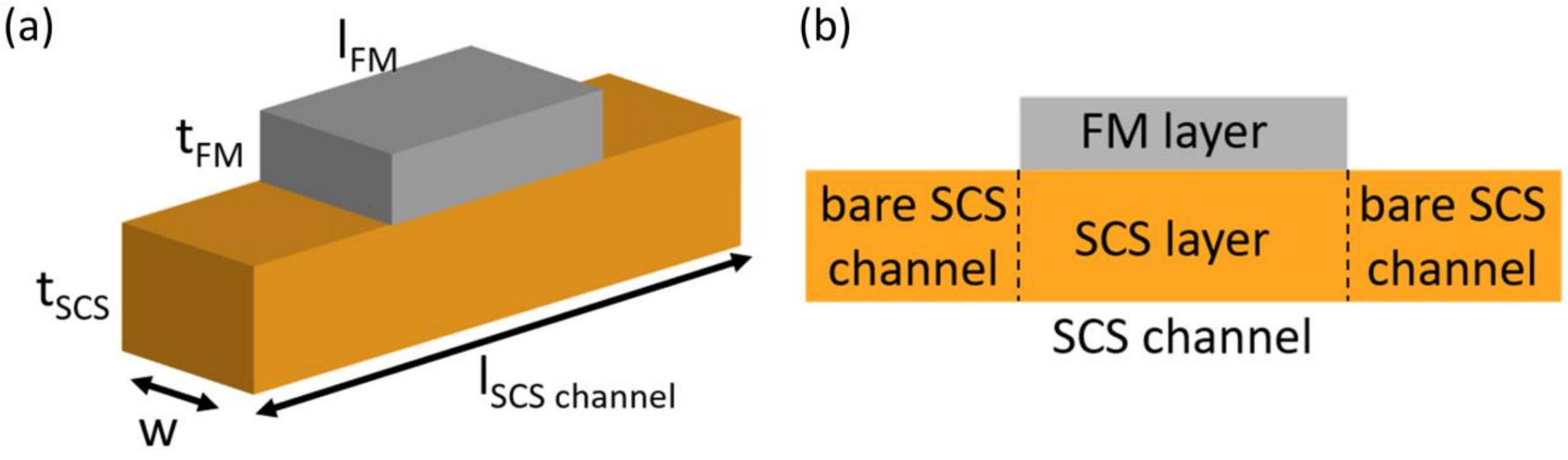

**Figure S2.** Schematic illustration of (a) a SOT device with an extending SCS channel and an above FM layer. (b) definition of bare channel, SCS layer and FM layer.

We also calculate the power consumption by damping-like SOT efficiency ($\xi_{DL}$) of SCS. To simplify the scenario, we only consider the power consumed by an extending SCS channel with an above FM layer under recent SOT-MRAM architecture, and exclude the contribution from the transistors and electrodes for driving memory elements, as shown in **Figure S2(a)**. Note that the entire SCS channel is partially covered by FM layer, here the (un)covered part is denoted as the bare SCS channel/SCS layer respectively, as shown in **Figure S2(b)**. The total write power consumption therefore can be estimated by

$$P = P_{\text{bare channel}} + P_{\text{SCS/FM}} = I_{total}^2 R_{\text{bare channel}} + I_{\text{SCS}}^2 R_{\text{SCS}} + I_{\text{FM}}^2 R_{\text{FM}}, \tag{S4}$$

where $P$, $P_{\text{bare channel}}$, and $P_{\text{SCS/FM}}$ respectively are the total write power consumption of whole SOT-MRAM element, the bare SCS channel, and the SCS/FM double layer; $I_{total}$, $I_{\text{SCS}}$ and $I_{\text{FM}}$ respectively are the charge current amount flowing in the bare SCS channel, in the SCS layer, and in the FM layer; $R_{\text{bare channel}}$, $R_{\text{SCS}}$ and $R_{\text{FM}}$ respectively are the resistance of the bare SCS channel, the SCS layer and the FM layer. After rearranging the equation, the critical power consumption per single bit can be written as

$$\begin{aligned} P = {} & [(1+s)J_c^{\text{SCS}} w t_{\text{SCS}}]^2 \rho_{\text{SCS}} \frac{l_{\text{scs channel}} - l_{\text{FM}}}{w t_{\text{scs}}} + (J_c^{\text{SCS}} w t_{\text{SCS}})^2 \rho_{\text{SCS}} \frac{l_{\text{FM}}}{w t_{\text{scs}}} \\ & + [s J_c^{\text{SCS}} w t_{\text{SCS}}]^2 \rho_{\text{FM}} \frac{l_{FM}}{w t_{\text{FM}}}, \end{aligned} \tag{S3}$$

where $s \equiv I_{\text{FM}}/I_{\text{SCS}} = t_{\text{FM}}\rho_{\text{SCS}}/t_{\text{SCS}}\rho_{\text{FM}}$; $J_c^{\text{SCS}}$ is the critical charge current density flowing in the SCS layer; $w$, $t_{\text{SCS(FM)}}$, $l_{\text{SCS channel(FM)}}$ respectively are the width of the device, thickness/length of total SCS channel and FM layer. Then,

$$\begin{aligned} P &= (1+s)^2 \rho_{\text{SCS}} {J_c^{\text{SCS}}}^2 V_{\text{bare channel}} + (1+s)\rho_{\text{SCS}} {J_c^{\text{SCS}}}^2 V_{\text{SCS}} \\ &= (1+s)^2 p V_{\text{bare channel}} + (1+s) p V_{\text{SCS}}, \end{aligned} \tag{S3}$$

where $V_{\text{bare channel}} = wt_{\text{SCS}}(l_{\text{scs channel}} - l_{\text{FM}})$, $V_{\text{SCS}} = wt_{\text{SCS}}l_{\text{FM}}$ respectively are the volume of the bare SCS channel and the SCS layer; $p = \rho_{\text{SCS}}{J_c^{\text{SCS}}}^2$ is the power consumption per single bit per volume without current shunting. To further exclude the geometry and FM materials differences between devices, the total power consumption can be represented as the following equation based on a macrospin model,

$$P = (1+s)^2C^2\rho_{\text{SCS}}{\xi_{DL}}^{-2}V_{\text{bare channel}} + (1+s)C^2\rho_{\text{SCS}}{\xi_{DL}}^{-2}V_{\text{SCS}}, \tag{S3}$$

which is from $J_c^{\text{SCS}} = \frac{C}{\xi_{DL}}$, and C depends on the SOT devices is with PMA[S11, 12] or IMA[S13-15]: $C_{PMA} = \left(\frac{2}{\pi}\right)\left(\frac{2e}{\hbar}\right)\mu_0 M_s t_{\text{FM}} H_c$, and $C_{IMA} = \left(\frac{2e}{\hbar}\right)\mu_0 M_s t_{\text{FM}}\alpha\left(H_c + \frac{M_{eff}}{2}\right)$. Where $e$ and $\hbar$ are the elementary charge and the reduced Plank constant, respectively; where $\mu_0$ is the vacuum permeability; $M_s$, $t_{\text{FM}}$, $H_c$, $\alpha$, and $M_{eff}$ are the saturation magnetization, thickness, coercivity, magnetic damping constant and the effective magnetization of the FM layer, respectively. To estimate the power consumption among various SCSs, we choose CoFeB as FM materials with PMA ($\rho_{\text{CoFeB}} = 190\ \mu\Omega\cdot\text{cm}$, $M_s = 1500\ emu/cm^3$, $H_c = 100\ Oe$), and the SOT device geometry is $w = l_{\text{FM}} = 0.2\ \mu m$, $l_{\text{SCS channel}}^{short} = 0.3\ \mu m$, $l_{\text{SCS channel}}^{long} = 0.6\ \mu m$, $t_{\text{SCS}} = 5\ nm$, $t_{\text{FM}} = 1\ nm$. The summary of calculated power consumption based on **Equation S3** is shown in **Figure S3** and **Table S2**. Notably, due to $\xi_{DL}$ of these SCSs are determined by different methods and the chosen FM materials and geometry are also inconsistent, the trend between **Figure S3** and **Figure 5** in main text shows a little difference. However, the relatively low resistivity and large damping-like SOT efficiency of $Pt_{0.57}Cu_{0.43}$ makes it still performs the lowest power consumption among Pt-based SCSs.

For the devices with extreme SCS channel length, the power consumption of a SOT device is dominated either by $P_{\text{bare channel}}$ or $P_{\text{SCS/FM}}$: $(1+s)V_{\text{bare channel}} > V_{\text{SCS}}$ for long SCS channel scenario, $(1+s)V_{\text{bare channel}} < V_{\text{SCS}}$ for short SCS channel scenario. One can conveniently compare the power consumption among SCSs with given device geometry and FM material by $P_{\text{bare channel}} \propto (1+s)^2\rho_{\text{SCS}}\xi_{DL}{}^{-2}$ or $P_{\text{SCS/FM}} \propto (1+s)\rho_{\text{SCS}}\xi_{DL}{}^{-2}$. However, the consumed power in real applications might be between short/long SCS channel scenario based on the geometry of SOT devices. Moreover, $H_c$, $\alpha$, $\xi_{DL}$ actually are affected by the SCS layer, both the SCS type and the interface condition might change the value of these parameters. Therefore, the fairest way to judge the performance among SCSs still is to measure the critical switching current density of real SOT devices with the same FM material and geometry.

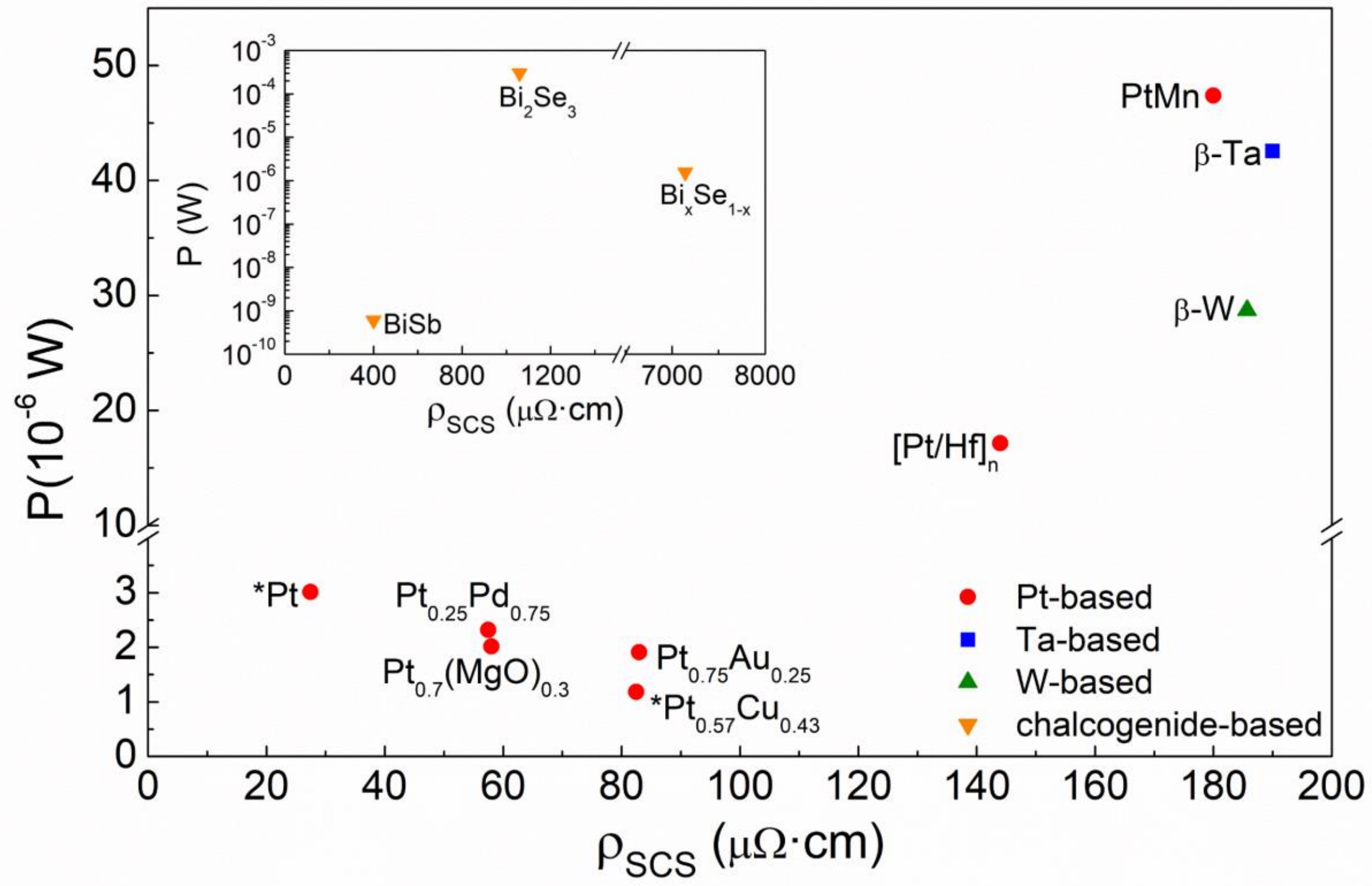

**Figure S3. Calculated switching power consumption.** The summary of power consumption ($P$) for current-induced SOT switching versus SCS resistivity ( $\rho_{\mathrm{SCS}}$ ) for a SOT device (SCS(5)/CoFeB(1)) with PMA and a short SCS channel ($l_{\mathrm{SCS\ channel}}^{short} = 0.3\ \mu m$). The estimation is based on a macrospin model and using reported damping-like SOT efficiency ($\xi_{DL}$) and $\rho_{\mathrm{SCS}}$.

| SCS type | SCS | $\rho_{\mathrm{SCS}}$ [μΩ·cm] | $\xi_{DL}$ [a.u.] | $J_c^{\mathrm{SCS}}$ [A·cm$^{-2}$] | $P^{short}$ [W] | $P^{long}$ [W] | Reference |
|---|---|---|---|---|---|---|---|
| Pt-based | $Pt_{0.57}Cu_{0.43}$ | 82.5 | 0.44 | $6.54 \times 10^{-6}$ | $4.17 \times 10^{-7}$ | $1.67 \times 10^{-6}$ | This work |
| Pt-based | Pt | 27.4 | 0.16 | $1.88 \times 10^{-7}$ | $1.02 \times 10^{-6}$ | $4.09 \times 10^{-6}$ | This work |
| Pt-based | PtMn | 180 | 0.11 | $2.63 \times 10^{-7}$ | $1.76 \times 10^{-5}$ | $7.07 \times 10^{-5}$ | [S1] |
| Pt-based | $Au_{0.25}Pt_{0.75}$ | 83 | 0.35 | $8.28 \times 10^{-6}$ | $6.72 \times 10^{-7}$ | $2.69 \times 10^{-6}$ | [S2] |
| Pt-based | $Pt_{0.25}Pd_{0.75}$ | 57.5 | 0.26 | $1.11 \times 10^{-7}$ | $8.03 \times 10^{-7}$ | $3.21 \times 10^{-6}$ | [S3] |
| Pt-based | $Pt_{0.7}(MgO)_{0.3}$ | 58 | 0.28 | $1.03 \times 10^{-7}$ | $6.99 \times 10^{-7}$ | $2.80 \times 10^{-6}$ | [S4] |
| Pt-based | $[Pt/Hf]_n$ | 144 | 0.16 | $1.81 \times 10^{-7}$ | $6.26 \times 10^{-6}$ | $2.50 \times 10^{-5}$ | [S5] |
| Ta-based | β-Ta | 190 | 0.12 | $2.41 \times 10^{-7}$ | $1.59 \times 10^{-5}$ | $6.38 \times 10^{-5}$ | [S6] |
| W-based | β-W | 185.7 | 0.14 | $2.01 \times 10^{-7}$ | $1.07 \times 10^{-5}$ | $4.30 \times 10^{-5}$ | [S7] |
| chalco.-based | $Bi_2Se_3$ | 1060 | 0.16 | $1.81 \times 10^{-7}$ | $1.56 \times 10^{-4}$ | $6.22 \times 10^{-4}$ | [S8] |
| chalco.-based | $Bi_xSe_{1-x}$ | 7143 | 18.6 | $1.56 \times 10^{-5}$ | $1.26 \times 10^{-6}$ | $5.02 \times 10^{-6}$ | [S9] |
| chalco.-based | BiSb | 400 | 52 | $5.58 \times 10^{-4}$ | $2.51 \times 10^{-10}$ | $1.00 \times 10^{-9}$ | [S10] |

**Table S2.** Summary of $\rho_{\mathrm{SCS}}$, $\xi_{DL}$ and calculated $J_c^{\mathrm{SCS}}$, $P$ of reported spin current sources based on macrospin model. $P^{short}$ and $P^{long}$ are the total calculated power consumption under different SCS channel length ($l_{\mathrm{SCS\ channel}}^{short} = 0.3\ \mu m$, $l_{\mathrm{SCS\ channel}}^{long} = 0.6\ \mu m$), including both the contribution from bare SCS channel and SCS/FM double layer.

### 3. A criterion for energy dissipation limitation

Since ideally only the current flowing in the SCS channel can contribute to SOT switching, current shunting into the adjacent FM layer results in a certain degree of energy dissipation. A criterion to evaluate how much dissipated energy can be caused by employing different SCSs therefore is crucial in designing realistic SOT memory devices. To further take the current shunting into account, we define another two engineering parameters: the geometry ratio $r_G = \frac{t_{\mathrm{FM}}}{t_{\mathrm{SCS}}}$ and the resistivity ratio $r_R = \frac{\rho_{\mathrm{FM}}}{\rho_{\mathrm{SCS}}}$, and then rearrange the Equation S3 into

$$p_{write} = \left[\left(1 + \frac{r_G}{r_R}\right)\frac{1}{1 + r_G}\right]\rho_{\mathrm{SCS}}{J_c^{\mathrm{SCS}}}^2 = (1+\eta)\rho_{\mathrm{SCS}}{J_c^{\mathrm{SCS}}}^2 = (1+\eta)p, \qquad \text{(S4)}$$

where $\eta$ is the energy dissipation proportion (dimensionless) due to current shunting, which can be evaluated by $r_G$ and $r_R$; $p = \rho_{\mathrm{SCS}}{J_c^{\mathrm{SCS}}}^2$ can be denoted as the power consumption per bit per unit volume without current shunting effect. For a certain FM materials system with given $r_G$, the maximum $\rho_{\mathrm{SCS}}$ to satisfy a certain $\eta$ limitation should be

$$\rho_{\mathrm{SCS}}^{max} = \left[1 + \left(1 + \frac{1}{r_G}\right)\eta\right]\rho_{\mathrm{FM}}. \qquad \text{(S5)}$$

The estimated $\rho_{\mathrm{SCS}}^{max}$ based on several common stacking geometry and FM materials are summarized in **Table S3**.

Take SCS(5)/CoFeB(1) heterostructure for example, if the $\eta$ limitation is set to be 50% (an extra $0.5p$ is required to achieve writing a single bit), then the resistivity of the SCS material should not exceed 760 μΩ·cm. If the $\eta$ limitation is set to be 10%, then the SCS resistivity should not exceed 304 μΩ·cm.

| $r_G$ | $\eta$ | $\rho_{SCS}^{max}/\rho_{FM}$ | $\rho_{SCS}^{max}$ for $\rho_{Co}$=26.6 μΩ·cm [μΩ·cm] | $\rho_{SCS}^{max}$ for $\rho_{CoFeB}$=190 μΩ·cm [μΩ·cm] |
|---|---|---|---|---|
| **1/2** | 0.1 | 1.3 | 35 | 247 |
| | 0.2 | 1.6 | 43 | 304 |
| | 0.5 | 2.5 | 67 | 475 |
| | 9 | 27 | 718 | 5130 |
| **1/3** | 0.1 | 1.4 | 37 | 266 |
| | 0.2 | 1.8 | 48 | 342 |
| | 0.5 | 3 | 80 | 570 |
| | 9 | 36 | 958 | 6840 |
| **1/5** | 0.1 | 1.6 | 43 | 304 |
| | 0.2 | 2.2 | 59 | 418 |
| | 0.5 | 4 | 106 | 760 |
| | 9 | 54 | 1436 | 10260 |

**Table S3.** Summary of $r_G$, $\eta$, $\rho_{SCS}^{max}/\rho_{FM}$, and the estimated $\rho_{SCS}^{max}$ under different $r_G$, $\eta$ limitation and adjacent FM layer.